\documentclass[11pt,twoside,usenatbib,useAMS]{article}

\usepackage{asp2006}
\usepackage{epsf}
\usepackage{psfig}
\usepackage{lscape}
\usepackage{natbib,graphicx}

\def\grtsim{\mathrel{\hbox{\rlap{\hbox{\lower2pt\hbox{$\sim$}}}\raise2pt\hbox{$>$}}}} 
\def\lesssim{\mathrel{\hbox{\rlap{\hbox{\lower2pt\hbox{$\sim$}}}\raise2pt\hbox{$<$}}}}

\def\degree{\nobreak\ifmmode{^\circ}\else{$^\circ$}\fi} 

\newcommand{\umu}{$\mu$}

\newcommand{\whzsr}{W\,Hz$^{-1}$\,sr$^{-1}$}

\newcommand{\lya}{Ly\,$\alpha$}

\newcommand{\nh}{$N_{\rm H}$}
\newcommand{\av}{$A_{\rm V}$}
\newcommand{\lbol}{$L_{\rm bol}$}

\newcommand{\ztwo}{$z\sim$2}
\newcommand{\zphot}{$z_{\rm phot}$}
\newcommand{\zspec}{$z_{\rm spec}$}

\begin{document}
\title{High-redshift obscured quasars}   %%% Fill in title
\author{Alejo Mart\'\i nez-Sansigre$^{1,2,\dag}$ and Steve Rawlings$^{1}$}   %%% Fill in author
                                %%% names

\affil{$^{1}$Astrophysics, Department of Physics, University of Oxford, Keble Rd,
  Oxford, OX1 3RH, UK}    %%% Fill in author affiliations

\affil{$^{2}$Max-Planck Insitut f\"ur Astronomie, K\"onigstuhl-17, D-69117, Germany}    %%% Fill in author affiliations
\affil{{\rm {\footnotesize$^{\dag}$E-mail: martinez@mpia-hd.mpg.de (AMS)} }}

\begin{abstract} %%% Abstract to run on from here.
  Using mid-infrared and radio criteria, we select a sample of candidate
  $z\sim2$ obscured quasars. Optical spectroscopy confirms about half of these
  as type-2 quasars, and modelling the population suggests 50-80\% of the
  quasars are obscured. We find some flat radio spectrum type-2
  quasars, and tentative evidence for obscuration unrelated to the torus.
  Using a similar sample, we also find evidence for a significant fraction of
  Compton-thick quasars.
\end{abstract}

%%% MAIN BODY OF TEXT GOES HERE. CONSULT "INSTRUCTIONS FOR AUTHORS USING
%%% LATEX2E MARKUP", SECTIONS 2.3-2.6 FOR HELP WITH EQUATIONS, FIGURES,
%%% AND TABLES.

\section{Introduction}

The population of high-redshift obscured (type-2) quasars (active galactic
nuclei with \lbol $\grtsim 10^{38}$ W) has proven an elusive one, despite
great progress in deep surveys. In radio-selected samples, the fraction
of unobscured (type-1) quasars to the total number of quasars (the quasar
fraction) is found to be $\sim0.5$ \citep {2000MNRAS.316..449W}, corresponding
to a ratio of type-2 to type-1 $\sim$1:1. At X-ray wavelengths, surveys have
been successful in finding large numbers of type-2 AGN, but the exact fraction
of type-2 quasars is still unclear \citep [for recent results, see
e.g. ][]{2006ApJ...646L.103W,2006MNRAS.tmp.1122D,2006A&AGilli}.

A possible explanation for the discrepancies is that some quasars are very
heavily obscured (with absorbing columns, \nh $\grtsim 10^{28}$ m$^{-2}$) and
are therefore very difficult to detect in X-ray surveys. An alternative
approach is to use the characteristic warm dust ($T\sim$ 100-1000 K)
signature of AGN to select obscured quasars based on their mid-infrared
properties, and indeed \citet {2006:lacy} confirmed the success of a 3.6 to 8
\umu m colour selection to find AGN (obscured and unobscured) at intermediate
redshifts.

Of particular cosmological relevance is the quasar fraction at \ztwo, around
the maximum in the quasar activity. We therefore devised a set of strict
selection criteria to look for \ztwo\, type-2 quasars. This has also practical
advantages, since above $z=1.7$ \lya~is visible with ground-based optical
spectrographs.

\section{Selection}

Quasars are bright at mid-infrared wavelengths due to the presence of warm
dust. At \ztwo, observed 24 \umu m corresponds to rest-frame $\sim8$ \umu m,
or dust at $\sim 360$ K. At these wavelengths, the obscuration
due to dust is negligible, unless the \av\, is extreme. The extinction due to
dust increases rapidly with decreasing wavelength, so that type-2 quasars that
are bright at 24 \umu m should be significantly fainter at 3.6 \umu m
(rest-frame $\sim$1.2 \umu m). In addition, if the quasar has a value of
\av$\grtsim10$, the 3.6-\umu m flux density will be dominated by the host
galaxy, instead of the quasar light. Since quasar host galaxies are typically
massive ellipticals, one can assume a typical galaxy mass, and use the
3.6-\umu m flux density to obtain a crude photometric redshift.

Therefore we demand objects to meet the following criteria: $S_{24 ~
  \mu \rm m} > 300~\mu$Jy and $S_{3.6~ \mu \rm m} \leq
45~\mu$Jy. Assuming a typical quasar SED, the 24-\umu m criterion
corresponds to slightly below the break in the quasar luminosity
function at $z=2$, while the $S_{3.6 \mu \rm m}$ flux density cut
corresponds to \zphot$\geq1.4$. This allows some margin of error for
our crude \zphot~cut.

These two criteria should select quasars with \av$\grtsim5$ and
\zphot$\grtsim1.4$. However, they are also likely to include a large number of
ultra-luminous infrared galaxies (ULIRGs) at $z\lesssim$1. We therefore
included a radio criterion: 350~$\mu$Jy $\leq$ $S_{1.4 \rm~GHz}$ $\leq$~2 mJy.
Only ULIRGs with very high star-formation rates ($\grtsim$ 100-350 $M_{\odot}$
yr$^{-1}$ of massive stars) are expected to make this cut, which therefore
minimises the contamination in our sample. The trade-off is that the quasars
we select are radio-intermediate ($L_{1.4\, \rm GHz}$$\sim10^{24}$ \whzsr), so
although their (rest-frame) 8-$\mu$m luminosity densities correspond to below
the break in the luminosity function, their radio properties make them more
rare than the radio-quiet quasars.

\section {The quasar fraction}

Applying this criteria, we found a number of candidates (21 in the 4 deg$^{2}$
of the Spitzer extragalactic First Look Survey, FLS), and optical
spectroscopy confirmed about half of these as narrow-line type-2 quasars at $z\geq
1.7$. One object at $z=1$ had no high-ionisation lines, and could plausibly be
a ULIRG contaminant. The rest of the objects had completely blank
spectra, and there is no contamination from type-1 quasars \citep[see][]{2006MNRAS.tmp..691M}.

\begin{figure}%[!h] 
\begin{center} 
\hspace{0.25cm}
\psfig{file=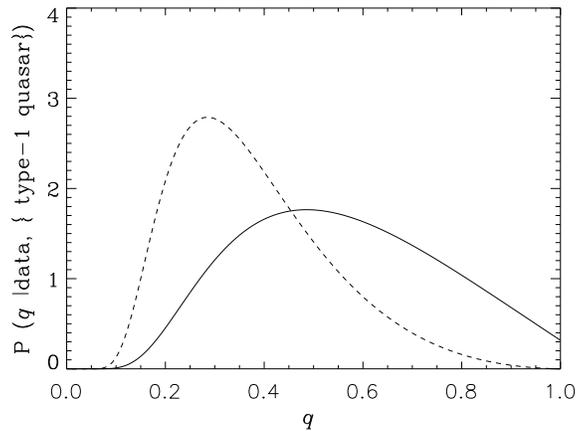,width=8cm,angle=90}  
\caption{\noindent Normalised posterior probability distribution of the quasar
  fraction, given the data and the predicted number of $z\geq2.0$ type-1
  quasars matched in 24-\umu m and radio criteria. The solid line shows the
  distribution if only the narrow line type-2 quasars at \zspec$\geq2.0$ are
  taken into acount, while the dashed line shows the distribution if the blank
  objects with \zphot$\geq2.0$ are also included.  
}
\label{fig:figure} 
\end{center} 
\end{figure}

To estimate the quasar fraction, we modelled the expected number of type-1
quasars meeting our 24-\umu m and 1.4 GHz criteria, and at $z\geq2.0$, where
our completeness is expected to be high.  Assuming a flat prior probability
distribution (PDF) for the quasar fraction (reflecting our ignorance), we
calculated the posterior PDF of the quasar fraction, given the predicted
number of type-1 quasars, and the number of type-2 quasars showing
narrow-lines and with \zspec $\geq2.0$. This is shown as the solid curve in
Figure~\ref{fig:figure}. The distribution peaks around a value of the quasar
fraction, $q$, $\sim0.5$ implying a similar number of narrow-line type-2
quasars as type-1 quasars.

Some of the blank objects will be in the ``spectroscopic desert'' (1.2 $\leq z
\leq$ 1.7), yet many are probably at $z\geq1.7$ \citep [and two cases have been
spectroscopically confirmed to be at $z=1.75$ and 2.25 from Spitzer spectroscopy by][]
{2006ApJ...638..613W}. 
We therefore use the
crude photometric estimates (from the 3.6 \umu m flux density) to estimate how
many blank-spectrum objects lie at $z\geq 2.0$. When these objects are added
to the narrow-line objects, one obtains the dashed-line distribution of
Figure~~\ref{fig:figure}, and the modal value for $q$ is now closer to 0.25. The true quasar
fraction is probably somewhere between $q=0.2$ and $q=0.5$, so that we can
conclude that at the peak of type-1 quasar activity, 50-80\% of quasars are
obscured \citep{2005Natur.436..666M}. 

A possible explanation for the blank objects is that they are being obscured
by dust on scales larger than the narrow-line region. The obscuration would
therefore be independent of the orientation.
If this scheme is true, then some of the
blank-spectrum objects are expected to harbor quasars with the torus face-on:
objects that would be type-1s if they were not obscured by dust on kpc scales.
If the radio jet is close enough to our line of sight, such a quasar might
exhibit a flat-spectrum. 

To test this, we used radio data at 1.4 and 4.9 GHz from the VLA and at 610
MHz from the GMRT to measure the radio spectral indices \citep
{2006MNRAS.tmpL.109M}. Most objects show steep spectra, as expected from radio
sources, and some gigahertz-peaked sources are also found. However, four
objects show flat radio spectra. One of these is the $z=1$ object, and its
radio luminosity ($L_{1.4 \rm GHz}$$\sim10^{23}$ \whzsr) is consistent with a
massive-star-formation rate of $\sim$300 $M_{\odot}$ yr$^{-1}$: it is probably
a ULIRG contaminant. The other three objects are probably flat-spectrum type-2
quasars, and none show narrow-emission lines (one of them has \zspec $=2.25$
from Spitzer spectroscopy).

\section{Compton-thick quasars}

If this scheme is correct, as suggested by the evidence collected so far, then
type-2 quasars outnumber type-1 quasars by $\sim$2-3:1, and therefore a
fraction of type-2 quasars seems to have been missed out by hard X-ray
surveys. The most likely explanation is therefore that some of these type-2
quasars are optically-thick to Compton-scattering, i.e. are ``Compton-thick''
(\nh $\geq10^{28}$ m$^{-2}$). Since no X-ray data was available for the first
sample, we selected a similar sample in the Subaru-XMM Deep Fiels (SXDF) where
deep X-ray data was available. The area here is smaller (0.8 deg$^{2}$), so we
needed to change our selection slightly to yield more quasars. We lowered the
lower cut in radio flux density to $S_{1.4 \rm GHz}\geq 100$ \umu Jy and left
all other criteria unchanged. As well as increasing the number of
high-redshift type-2 quasars, this would increase the number of ULIRG
contaminants, and no spectroscopy was originally available to distinguish
between objects.

To separate the type-2 quasars from the ULIRGs, we fitted the spectral energy
distributions (SEDs) between B-band and 24 \umu m,  using a quasar and a ULIRG model, and used a bayesian
criterion (the odds ratio) to select between models. We only kept the
objects where the quasar model was strongly favoured over the ULIRG model, with reasonably good
fits, and with \zphot $\geq 1.70$, so that they could be tested
observationally. This left us with 12 good type-2 candidates, only two of
which were detected in the deep X-ray catalogue of the SXDF. From the
photometric redshifts and the X-ray data, the two objects have estimated
column densities of log$_{10}(N_{\rm H} / \rm m^{2})=$ 27.25 and 27.50, so they
are both heavily obscured but Compton-thin.

The other 10 objects are undetected in X-rays, up to an approximate flux limit
in the 2-12 keV band $S_{2-12 \rm keV}$$\sim$3$\times10^{-18}$ W m$^{-2}$.
They have similar estimated values of \zphot and bolometric luminosities to
the two X-ray detections, so if they are indeed quasars, they are probably
more heavily absorbed, with \nh$\grtsim10^{28}$ m$^{-2}$. We therefore find
evidence that many of these objects are Compton-thick (Mart\'\i nez-Sansigre
et al. 2006c).

\acknowledgements %%% Text of acknowledgements runs on after this command.

We thank our  collaborators for their important contributions to this work.

\end{document}